\documentclass[prb,twocolumn,showpacs,amsmath,amssymb]{revtex4}
\usepackage{graphicx}
\usepackage{dcolumn}
\usepackage{bm}
\usepackage{color}

\usepackage{float,afterpage,wrapfig}
\usepackage[caption=false]{subfig}
\usepackage{multirow}
\usepackage[normalem]{ulem} 
\usepackage{placeins}

\DeclareMathAlphabet{\bi}{OML}{cmm}{b}{it}
\newcommand{\bk}{\mathbf{k}}
\newcommand{\dn}{\downarrow}
\newcommand{\half}{\text{$\textstyle\frac{1}{2}$}}
\newcommand{\pdag}{\phantom{\dag}}
\newcommand{\sgn}{\operatorname{sgn}}
\newcommand{\up}{\uparrow}

\begin{document}
\title{Critical local moment fluctuations and enhanced pairing correlations
in a cluster Anderson model}
\author{Ang Cai}
\affiliation{Department of Physics and Astronomy, Rice Center for Quantum Materials, 
Rice University,
Houston, Texas, 77005, USA}
\author{J.\ H.\ Pixley}
\affiliation{Department of Physics and Astronomy, Center for Materials Theory, Rutgers University, 
Piscataway, NJ 08854, USA}
\affiliation{Condensed Matter Theory Center and the Joint Quantum Institute,
Department of Physics, University of Maryland,
College Park, Maryland 20742-4111, USA}
\author{Kevin Ingersent}
\affiliation{Department of Physics, University of Florida, Gainesville, Florida
32611-8440, USA}
\author{Qimiao Si}
\affiliation{Department of Physics and Astronomy, Rice Center for Quantum Materials,
Rice University,
Houston, Texas, 77005, USA}

\date{\today}
\begin{abstract}
The appearance of unconventional superconductivity near 
heavy-fermion quantum critical points (QCPs) motivates investigation
of pairing correlations close to a ``beyond Landau'' Kondo-destruction
QCP. We focus on a two-Anderson-impurity cluster in which
Kondo destruction is induced by a pseudogap in the conduction-electron
density of states. Analysis via continuous-time
quantum Monte-Carlo and the numerical renormalization
group reveals
a previously unstudied QCP that both displays the critical-local
moment fluctuations characteristic of
Kondo destruction and leads to a strongly enhanced singlet-pairing susceptibility.
Our results 
provide
new insights into the mechanism for 
 superconductivity in
quantum critical metals.

\end{abstract}

\maketitle
\section{Introduction}
Heavy-fermion metals are highly tunable and provide a prototype setting to explore
strong correlation physics in general \cite{LeeNagaosaWen, Si2016,Takabayashi2009}.
In heavy-fermion systems, unconventional superconductivity often develops near 
their quantum critical points (QCPs) \cite{SiSteglich_Science2010,Lohneysen_RMP2007}.  
Detailed theoretical and experimental studies have provided evidence for different classes 
of QCP. 
One class follows the Landau theory,
in which criticality is dictated by the fluctuations of an order parameter \cite{Hertz, Millis, Moriya}.
Another class of QCP goes beyond the Landau framework, in that it involves new
critical modes besides order-parameter fluctuations. 
The additional critical modes describe a critical destruction of the Kondo
entanglement between the localized magnetic moments and conduction electrons
\cite{Si-2001, Coleman-2001},
which is a form of electronic localization-delocalization instability.
As such, studies of superconducting pairing driven by Kondo-destruction quantum criticality 
elucidate unconventional superconductivity not only in heavy-fermion metals 
but also in a broad range of other correlated electron systems.

An important example of a Kondo-destruction QCP occurs in
CeRhIn$_5$, which has the highest $T_{c}$ among all the Ce-based 
heavy-fermion superconductors \cite{Park_Nature2006,Park_2008,Qimiao_JPSJ_1,
Qimiao_JPSJ_2} and is generally believed to have a $d_{x^{2}-y^{2}}$ 
pairing symmetry. 
A sudden change of the Fermi-surface size across the antiferromagnetic
QCP in CeRhIn$_5$, accompanied by
a diverging tendency of the carrier
effective mass \cite{Shishido.2005}, defy
explanation within the Landau-based
(spin-density-wave) scenario but instead provide evidence
supporting the Kondo-destruction
picture.

How unconventional superconductivity arises near a Kondo-destruction QCP
has yet to receive systematic theoretical study. The question
is challenging because the normal state is a non-Fermi liquid
with quasiparticles turned critical.
An avenue has been opened by the
development of a cluster extended dynamical mean-field
theory (C-EDMFT) \cite{Formalism}, which maps the periodic Anderson model
onto a cluster model coupled to self-consistently determined fermionic
and bosonic baths, where the latter decohere and eventually
destroy the Kondo entanglement \cite{entanglement}.
The Kondo destruction QCP of the lattice problem is embedded in the QCP of the quantum cluster model, 
and the multi-site cluster allows the development of unconventional pairing.

It it illuminating to study the quantum cluster model by itself. Previous studies
\cite{Ingersent.Si.2002,Glossop.Kirchner.Pixley.Si.2011,pixley2013quantum,
kirchner2008scaling,pixley2012kondo} demonstrate that quantum impurity models can
manifest hallmarks of a Kondo-destruction QCP such as a vanishing Kondo
energy scale, $\omega/T$ scaling of the dynamics, and a fractional exponent in the
temperature dependence of the local spin susceptibility. 
This is largely due to the fact that Kondo destruction is primarily a local phenomenon,
and the neglect of spatial correlation is relatively unimportant. 
Recent work in an Ising-anisotropic cluster Bose-Fermi Anderson model
\cite{Pixley.2015} has found enhanced pairing correlations near the Kondo-destruction QCP. 
This finding raises an important question: Does Kondo-destruction quantum criticality
robustly promote superconducting pairing correlations?

This work investigates a two-impurity pseudogap Anderson 
model with Ising exchange between the impurity spins. 
We show that the model exhibits a Kondo-destruction QCP that has not been discussed in
previous publications.
Instead of the coupling to a bosonic bath that was responsible for Kondo destruction
in Ref. \onlinecite{Pixley.2015}, here the driving force is exchange
coupling of the impurity spins to a conduction band with a
density of states that vanishes in power-law fashion at the
Fermi energy \cite{fradkin}.
The presence of a different mechanism for Kondo destruction allows us to address
the generality with which this type of quantum criticality promotes superconducting pairing.

\section{Model and solution methods}
The two-impurity Ising-anisotropic Anderson
Hamiltonian is
\begin{align}
\label{eqn:H}
H &= \sum_{\bk,\sigma} \epsilon_{\bk} \, c^{\dag}_{\bk\sigma}
     c^{\pdag}_{\bk\sigma} + \frac{V}{\sqrt{N_k}} \sum_{\bk,j,\sigma}
		 \bigl( e^{i\bk \cdot \textbf{r}_j} d^{\dag}_{j\sigma}
		  c^{\pdag}_{\bk\sigma} + H.c. \bigr) \notag \\
&+ \epsilon_{d} \sum_{j,\sigma} d^{\dag}_{j\sigma}
    d^{\pdag}_{j\sigma} + U \sum_{j} n_{j\up} n_{j\dn}
		+ I_{z}S_{1}^{z}S_{2}^{z},
\end{align}
with $\epsilon_{\bk}$ being the conduction-electron dispersion,
$V$ the hybridization (assumed to be local), $N_k$ the number of unit
cells in the host, $\epsilon_d$ the impurity level energy, $U$ the
on-site repulsion, and $I_z$ the Ising exchange coupling between
impurities at positions $\mathbf{r}_j$ $(j = 1$, $2$);
$n_{j\sigma} = d^{\dag}_{j\sigma} d^{\pdag}_{j\sigma}$ for
$\sigma = \,\uparrow$, $\downarrow$, and
$S_j^z=\half(n_{j\uparrow}-n_{j\downarrow})$.
The conduction-band density of states is chosen to be
\begin{equation}
\rho(\epsilon) = \frac{1}{N_k} \sum_{\bk} \delta(\epsilon -\epsilon_{\bk})
  = \rho_{0} |\epsilon / D|^r \, \Theta(D- |\epsilon|),
\label{eqn:DOS}
\end{equation}
where $D$ is the half-bandwidth. For $r>0$, $\rho(\epsilon)$ has a
pseudogap at the Fermi energy $(\epsilon=0)$.
The impurity-band coupling is
fixed
 by 
$\Gamma(\epsilon)= \pi \sum_{k} V^{2} \delta(\epsilon-\epsilon_{k})
= \Gamma_{0} |\epsilon/D|^{r}$, the hybridization function with
$\Gamma_{0} = \pi \rho_{0} V^{2}$.

For simplicity, we consider only the particle-hole-symmetric case
$\epsilon_d=-U/2$ and take the limit of infinite separation
$|\textbf{r}_{1}-\textbf{r}_{2}|$
in which there is a vanishing
hybridization-induced Ruderman-Kittel-Kasuya-Yosida interaction and the
two impurities are coupled only via the Ising exchange $I_z$.
With $I_z = 0$, we have two independent one-impurity pseudogap models;
for $0<r<\half$, a Kondo-destruction QCP \cite{Bulla.1997,Buxton.1998}
that we denote CR1 separates a Kondo phase ($\Gamma_0>\Gamma_c$) from a
local-moment phase ($\Gamma_0<\Gamma_c$). With $\Gamma_0=0$ and
$I_z>0$, the two impurity spins are decoupled from the conduction
band and anti-align in an Ising antiferromagnetic configuration.
Our goal is to probe the quantum phase transitions that arise when
both $\Gamma_0 > 0$ and $I_z > 0$ \cite{Garst.etal}.

\begin{figure}[t]
\captionsetup[subfigure]{labelformat=empty}
  \centering
 	\mbox{\includegraphics[width=1.0\columnwidth]{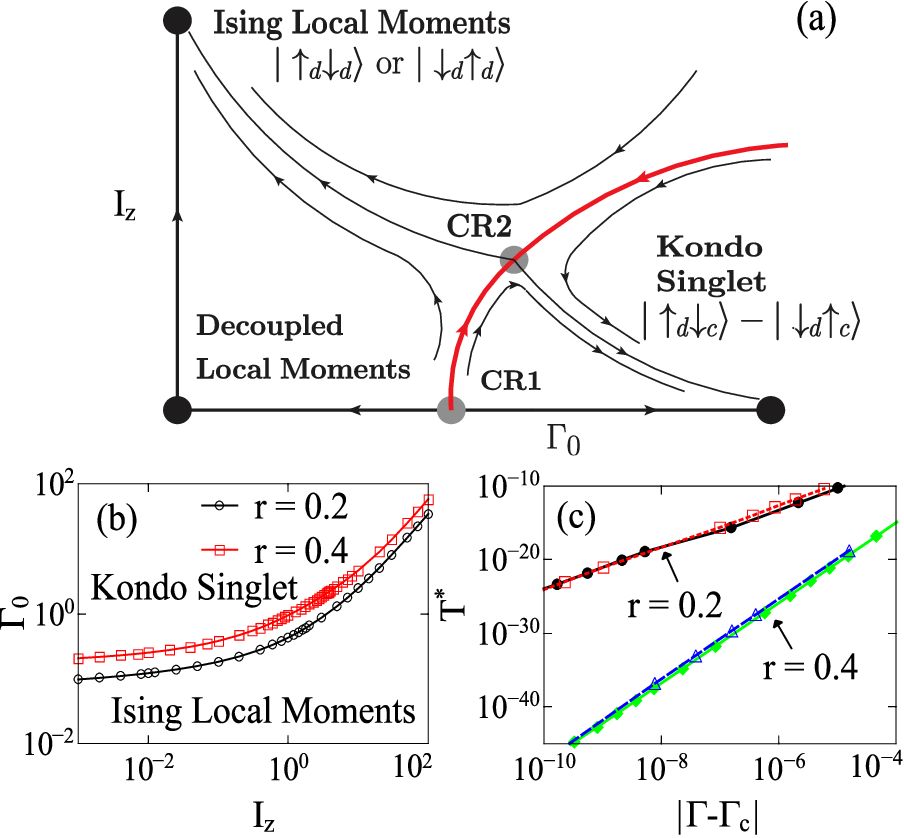}}
\\[-2ex]
\caption{\label{fig:1}
(a) Conjectured RG flow of the symmetric two-impurity pseudogap Anderson model.
Gray dots represent unstable fixed points and black dots represent stable fixed
points. CR1 is the unstable fixed point of the single-impurity pseudogap Anderson
model. CR2 is the unstable fixed point of the two-impurity model studied in this
work. The red line marks the separatrix and phase boundary.
(b) Phase boundary of the symmetric two-impurity pseudogap Anderson model on the
$I_z$--$\Gamma_0$ plane for $r=0.2$,
$U=-2\epsilon_d=0.3$ and for $r=0.4$,
$U=-2\epsilon_d=0.1$. The boundary value of $\Gamma_{0}$ obtained from NRG
calculations is plotted before extrapolation to the continuum limit (see discussion in Appendix \ref{sec:appendix1}).
(c) Crossover scale $T^{*}$ from the NRG vs $|\Gamma_0-\Gamma_c|$ on both
sides of the phase boundary for $r=0.2$, $I_{z}=1.54$, $\Gamma_c\simeq 0.5503$
and for $r=0.4$, $I_{z}=0.73$, $\Gamma_c\simeq 0.8032$. Filled symbols represent $\Gamma>\Gamma_{c}$ 
and open symbols represent $\Gamma<\Gamma_{c}$. 
Fits to
$T^{*}\propto|\Gamma_0-\Gamma_c|^{\nu}$ yield estimated exponents given
in the text.}
\end{figure}

We begin by analyzing the perturbative effect of the coupling
$I_z$ near the single-impurity critical point CR1. At this QCP,
$\langle S_{i}^{z}(\tau) S_{i}^{z} \rangle \sim \tau^{-(1-x_1)}$,
with $x_1$ being an $r$-dependent exponent that satisfies
$0<x_1(r)<1$ \cite{Ingersent.Si.2002}. Since the impurities decouple,
$\langle S_{1}^{z}(\tau)S_{2}^{z}(\tau) S_{1}^{z}S_{2}^{z} \rangle \sim
\tau^{-2(1-x_1)}$. The scaling dimension of $S_{1}^{z}S_{2}^{z}$
is thus seen to be $1-x_1(r)$ and we obtain the scaling dimension
$[I_{z}]=x_1(r)$. The Hamiltonian term $I_{z}S_{1}^{z}S_{2}^{z}$ is
therefore a relevant perturbation at CR1 and will likely lead the
two-impurity model to a new unstable fixed point CR2 as shown on a
conjectured RG flow diagram in Fig.\ \ref{fig:1}(a).
 
Since the pseudogap breaks conformal invariance, the model \eqref{eqn:H}
cannot be treated nonperturbatively using conventional analytical
methods \cite{Affleck.Ludwig,Gan,Zarand} and we instead employ continuous-time quantum Monte-Carlo
(CT-QMC) \cite{Werner-2007,Werner-2010} and the numerical renormalization
group (NRG) \cite{Wilson.1975,Bulla.2008}.
We present results for two representative cases: (i) $r=0.2$, $U=0.3$ and
(ii) $r=0.4$, $U=0.1$, where we have set the energy scale $D=1$.
In CT-QMC calculations we vary $I_{z}$ at fixed $\Gamma_{0}$, and are able to
reach sufficiently low temperatures to access the asymptotic quantum critical
regime. We
fix $I_z$ and vary $\Gamma_{0}$ when applying the
NRG, a technique that can reach arbitrarily close to absolute zero but
has limited ability to calculate finite-temperature dynamics.
Further numerical details
are described in Appendix \ref{sec:appendix1}. 

\section{Quantum critical properties}
A critical phase boundary can be mapped out within the NRG by looking
for the hybridization width $\Gamma_c(I_z)$ at which the asymptotic
low-energy many-body spectrum jumps from that of one stable fixed
point to another. The phase boundaries
are plotted Fig.\ \ref{fig:1}(b).
For $\Gamma_0$ close to $\Gamma_c$, the NRG spectrum
flows away from the critical spectrum toward one or other of the stable
fixed points around a crossover temperature
$T^{*} \propto |\Gamma_{0}-\Gamma_{c}|^{\nu}$. Using this relation,
illustrated in Fig.\ \ref{fig:1}(c), one obtains $\nu^{-1}=0.334(2)$ for
$r=0.2$ and $\nu^{-1}=0.1835(4)$ for $r=0.4$.

\begin{figure}[t]
\captionsetup[subfigure]{labelformat=empty}
  \centering
  \mbox{\includegraphics[width=1.0\columnwidth]{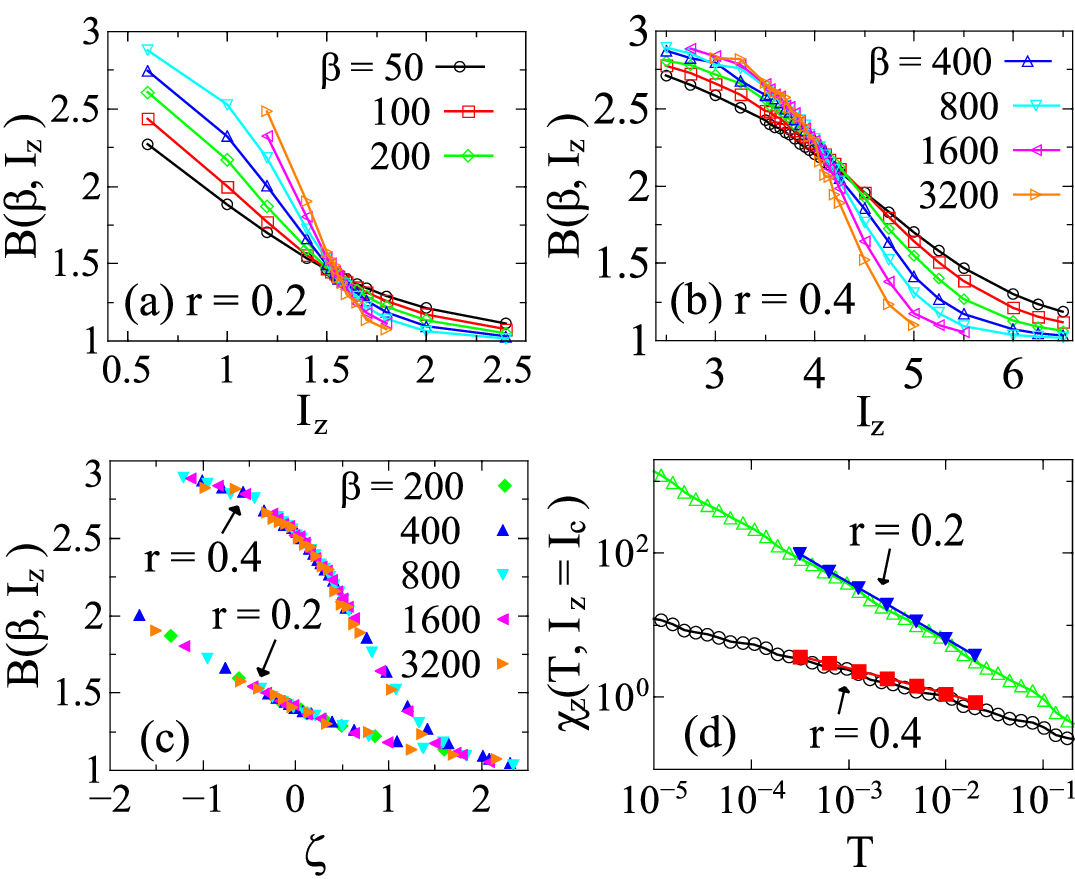}}
\\[-2ex]  
\caption{\label{fig:2} 
(a,b) Binder ratio $B(\beta, I_{z})$ from CT-QMC vs $I_{z}$ at various
inverse temperatures $\beta$ for
(a) $r=0.2$, $\Gamma_{0}=0.5$ and
(b) $r=0.4$,
$\Gamma_{0}=1.5$. (c) Scaling collapses of $B(\beta,I_{z})$ with respect to $\zeta= \beta^{1/\nu} (I_{z}-I_{c})/I_{c}
  + C\beta^{-\phi/\nu}$
giving $I_{c}=1.56(7)$, $\nu^{-1}=0.33(4)$ at $r=0.2$ and $I_{c}=3.75(7)$, $\nu^{-1}=0.20(2)$ at $r=0.4$.
(d) Static staggered local spin susceptibility $\chi_z$ vs $T$ at the estimated
location $I_z=I_c$ of the QCP, calculated using CT-QMC (filled symbols) and the NRG (open symbols). Fitting to
Eq.\ \eqref{chi_z_diverge} yields the values of $x$ given in the text.}
\end{figure}

To search for a 
QCP using CT-QMC, we examine the Binder ratio \cite{Binder}
$B(\beta, I_{z}) = \langle M^{4} \rangle / \langle M^{2} \rangle ^{2}$, 
where the staggered impurity magnetization
$M = \beta^{-1} \int_{0}^{\beta} d \tau \left[ S_{1}^{z}(\tau)
- S_{2}^{z}(\tau) \right]$. 
Plots of $B(\beta, I_{z})$ vs $I_z$ for different values of $\beta=1/k_B T$
should all cross at the location $I_{z}=I_{c}$ of a QCP, as is indeed
shown in Fig.\ \ref{fig:2}(a) for $r=0.2$ and Fig.\ \ref{fig:2}(b) for $r=0.4$.
A scaling collapse 
\begin{equation}
B(\beta, I_{z})= f \bigl( \beta^{1/\nu} (I_{z}-I_{c})/I_{c}
  + C\beta^{-\phi/\nu} \bigr)
\label{Binder_collapse}
\end{equation}
(where the term involving $C$ accounts for sub-leading finite temperature corrections)
demonstrates that the quantum phase transition at $I_{z}=I_{c}$ 
is second order, as illustrated in Fig.\ \ref{fig:2}(c). By minimizing
a quality function \cite{Houdayer.Hartmann.2004} (see Appendix \ref{sec:appendix2} for details),
we find
$\nu^{-1}=0.33(4)$ for $r=0.2$ and $\nu^{-1}=0.20(2)$ for $r=0.4$,
reproducing the NRG values  to within estimated
errors \cite{correction_to_scaling}.

The static staggered local spin susceptibility (the order-parameter
susceptibility),
defined as $\chi_{z} = \beta \langle M^{2} \rangle$,
diverges at the QCP as
\begin{equation}
\chi_{z}(I_{z}=I_{c}, T) \sim T^{-x},
\label{chi_z_diverge}
\end{equation}
as seen in Fig.\ \ref{fig:2}(d).
The values of $x(r)$ from CT-QMC [$x(0.2)=0.78(4)$ and $x(0.4)=0.34(5)$]
and the NRG [$x(0.2)=0.78588(3)$ and $x(0.4)=0.35075(3)$] are in good agreement.
We have also calculated the connected spin susceptibility,
$\chi^{c}_{z} = \beta(\langle M^{2}\rangle-\langle|M|\rangle^2)$,
which based on the scaling hypothesis can be described by
$\chi^{c}_{z}(\beta, I_{z}) = \beta^{x} g\left(\beta^{1/{\nu}}
(I_{z}-I_{c}\right)/I_{c} + C\beta^{-\phi/\nu})$;
see Fig.\ \ref{fig:connected} in Appendix \ref{sec:appendix3}. 

We summarize our results for the critical exponents $\nu^{-1}$ and $x$ 
at the two-impurity pseudogap QCP CR2
in Table \ref{table:1},
where we have also included NRG values of the order-parameter
critical exponent $\beta'$ defined through
$M(\Gamma_{0},T=0,h=0)\propto(\Gamma_{c}-\Gamma_{0})^{\beta'}$ and
the magnetic critical exponent $1/\delta$ defined through
$M(\Gamma_{0}=\Gamma_{c},T=0,h) \propto |h|^{1/\delta}$,
$h$ being
an external field that couples solely to the staggered impurity spin
(see Fig.\ \ref{fig:beta_delta} in Appendix \ref{sec:appendix4}). These exponents
take values different
from those at the single-impurity pseudogap QCP CR1 \cite{Ingersent.Si.2002}, 
demonstrating 
CR2 
to be a distinct critical
point. Moreover, they obey
scaling relations $\delta^{-1}=(1-x)/(1+x)$ and $\nu^{-1}=(1-x)/2\beta'$ 
characteristic of an interacting critical point \cite{Ingersent.Si.2002}.

\begin{table}[b]
\centering
\begin{tabular}{llllll}
\hline\hline
$r$   & source & $1/\nu$ & $x$        & $\beta'$    & $1/\delta$ \\ \hline
0.2 & CT-QMC & 0.33(4)    & 0.78(4)    &            &     \\
    & NRG    & 0.334(2)   & 0.78588(3) & 0.31991(2) & 0.11990(4)         \\
0.4 & CT-QMC & 0.20(2)    & 0.34(5)    &            &     \\
    & NRG    & 0.1835(4)  & 0.35075(3) & 1.7701(2)  & 0.48066(4)        \\ \hline\hline
\end{tabular}
\caption{\label{table:1}
Critical exponents (defined in the text) at the two-impurity pseudogap QCP CR2.
Parentheses enclose the estimated error in the last decimal place.}
\end{table}

We turn to the dynamical properties at CR2 of the single-particle Green's
function $G_{i,\sigma}(\tau,T) = \langle T_{\tau}
d_{i,\sigma}^{\dag}(\tau) \, d_{i,\sigma} \rangle$ and the spin correlation
function $\chi_{z}(\tau,T) = \langle T_{\tau} [S_{1}^{z}(\tau)\!-\!
S_{2}^{z}(\tau)] ( S_{1}^{z}\!-\!S_{2}^{z})\rangle$. Guided by previous work
on the single-impurity models \cite{Glossop.Kirchner.Pixley.Si.2011,
kirchner2008scaling, pixley2012kondo, pixley2013quantum}, we find
from CT-QMC (see Fig.\ \ref{fig:G_chi_critical} in Appendix \ref{sec:appendix3}) that these
functions share similar power-law forms in the low-$T$, large-$\tau$
limit:
\begin{eqnarray}
G_{i,\sigma}(\tau,T)&\sim & [ \pi T/ \sin (\pi \tau T) ]^{\eta_{G}(r)},
\\
\chi_{z}(\tau,T)&\sim & [ \pi T/ \sin (\pi \tau T) ]^{\eta_{\chi}(r)},
\label{w/T_scaling}
\end{eqnarray}
with
exponents $\eta_{G}(0.2)=0.795$, $\eta_{G}(0.4)=0.600$,
$\eta_{\chi}(0.2)=0.213$, and $\eta_{\chi}(0.4)=0.657$.
As expected, $\eta_{\chi}=1-x$ is satisfied within numerical accuracy.
Moreover, our results suggest that (i) the relation $\eta_{G}=1-r$
known to hold at CR1 \cite{Kircan} also applies
at CR2, and (ii) $0<\eta_{G}<1$ and $0<\eta_{\chi}<1$, so
$G_{i,\sigma}$ and $\chi_{z}$ will also obey $\omega/T$ scaling
on the real frequency axis \cite{Glossop.Kirchner.Pixley.Si.2011}.
This supports the 
interacting nature of CR2.

\section{Pairing susceptibilities}
\begin{figure}[t]
\captionsetup[subfigure]{labelformat=empty}
  \centering
  \mbox{\includegraphics[width=1.0\columnwidth]{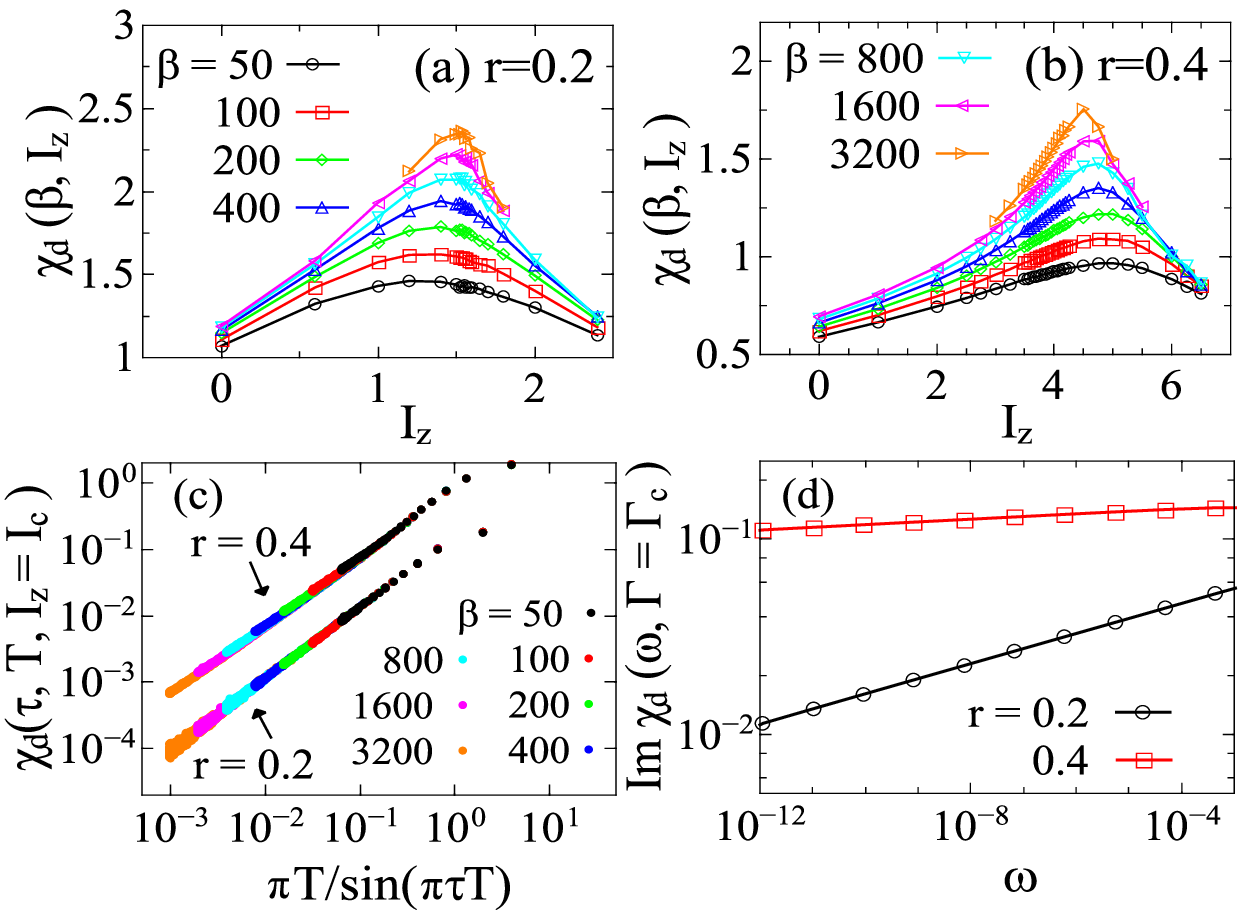}}                
  \\[-2ex]
\caption{\label{fig:3} 
Singlet pairing susceptibility: (a,b) Static susceptibility $\chi_{d}(\beta,I_{z})$
vs $I_{z}$ at various inverse temperatures $\beta$ for (a) $r=0.2$, $\Gamma_{0}=0.5$,
and (b) $r=0.4$, $\Gamma_{0}=1.5$.
(c) Imaginary-time susceptibility $\chi_{d}(\beta,\tau)$ at $I_{z}=I_{c}$,
consistent with a $1/\tau^{1+y}$ decay with $y=0.075$ for $r=0.2$, $\Gamma_0=0.5$
and $y=0.012$ for $r=0.4$, $\Gamma_0=1.5$. For clarity, the $r=0.4$ susceptibilities
have been multiplied by a factor of $10$.
(d) NRG results for the imaginary part of the real-frequency susceptibility,
$\operatorname{Im} \chi_{d}(\omega) \propto \omega^{y}$, at
$\Gamma_{0}=\Gamma_{c}$, $T=0$, calculated both for $r=0.2$, $I_{z}=1.54$
yielding $y=0.077(1)$ and for $r=0.4$, $I_{z}=0.73$ yielding $y=0.0139(1)$.}
\end{figure}

We study static pairing susceptibilities $\chi_{\alpha}(\beta,I_z) =
\int_{0}^{\beta} d\tau \langle T_{\tau} \Delta_{\alpha}^{\dagger}(\tau)
\, \Delta_{\alpha} \rangle$ with
$\Delta_d=(d_{2\dn}d_{1\up}-d_{2\up}d_{1\dn})/\sqrt{2}$ (singlet channel)
\cite{footnote}
and $\Delta_p=(d_{1\up}d_{2\up}+d_{1\dn}d_{2\dn})/\sqrt{2}$ (triplet channel).
Using the general four-point correlation function formula in CT-QMC
\cite{performance_analysis}, we find singlet pairing to be significantly
enhanced near the QCP, as shown in Figs.\ \ref{fig:3}(a) and
\ref{fig:3}(b). By contrast, triplet pairing
is monotonically suppressed as $I_{z}$ increases
(see Fig.\ \ref{fig:triplet} in Appendix \ref{sec:appendix3}).

At $T=0$, the imaginary part of the dynamical pairing susceptibility
$\operatorname{Im} \chi_{d}(\omega)$ can be calculated using the NRG.
We plot data for
$\Gamma_0=\Gamma_c(I_z)$ in Fig.\ \ref{fig:3}(d) and for other
cases in Fig.\ \ref{fig:Im_chi_Kondo_LM} in Appendix \ref{sec:appendix4}.
Our results can be summarized in the form
\begin{equation}
\operatorname{Im} \chi_d(\omega) \sgn(\omega) \propto
\begin{cases}
|\frac{\omega^{*}}{D}|^{y} |\frac{\omega}{\omega^{*}}| ^{1 -2r}   &  |\omega| < \omega^{*}, I_{z}<I_{c} \\
|\frac{\omega^{*}}{D}|^{y} |\frac{\omega}{\omega^{*}}| ^{1 +2r}   &  |\omega| < \omega^{*}, I_{z}>I_{c} \\
|\frac{\omega}{D}|^{y}  &   \omega^{*} < |\omega| < \omega_{1},
\end{cases}
\label{pair_freq}
\end{equation}
where $\omega_{1}$ is the high-energy scale marking the upper bound of the
quantum critical regime and $\omega^{*}\simeq T^{*}$ is the scale for
crossover into the low-temperature phase. This implies that at the
critical point, $\chi_{d}(\tau)\sim 1/\tau^{1+y}$, cf.\ Fig.\
\ref{fig:3}(c). The NRG gives $y=0.077(1)$ for $r=0.2$ and $y=0.0139(1)$
for $r=0.4$, values that agree very well with the CT-QMC estimates
of $y=0.075$ and $y=0.012$, respectively.
Equation \eqref{pair_freq} also implies (see Appendix \ref{sec:appendix5} for derivation)
that near the QCP,
\begin{align}
\operatorname{Re} \chi_{d}(\omega=0) =
& C_{1}(r) - C_{2}(r) \left( \frac{1}{y}- \frac{1}{ 1\pm 2r } \right)
\notag \\
&\times \left( \frac{|(I_{z}-I_{c})/I_{c}|^{\nu}}{D} \right) ^{y}
\label{eq:pair_peak}
\end{align}
with $\pm$ corresponding to $I_{z}>I_{c}$ or $I_{z}<I_{c}$,
and $C_{1}(r)$ and $C_{2}(r)$ being independent of $I_{z}$. Given that
$y\nu \ll 1$, $\operatorname{Re} \chi_{d}(\omega=0)$ should have a pronounced
cusp at $I_{z}=I_{c}$, as confirmed by the numerical data in Figs.\
\ref{fig:3}(a) and \ref{fig:3}(b).

\section{Discussion and Summary}
We note that, in the single-impurity
pseudogap Anderson model, the impurity spectral function vanishes (diverges) 
as $ |\omega|^{r}$ ($|\omega|^{-r}$) in the local-moment (Kondo-screened) phase \cite{Bulla.1997}.
Our calculations suggest that this property also holds in the present model 
(see Fig.\ \ref{fig:G_tau_Kondo_LM} in Appendix \ref{sec:appendix2}). This indicates that the frequency dependences of
$\operatorname{Im} \chi_d(\omega)$ at the two stable fixed points 
do not acquire any singular correction \cite{Buxton.1998}. 
By contrast, at CR2 the $|\omega|^{y}$ dependence reflects the relevance of vertex corrections at
an interacting critical point.
Since CR1 and CR2 both exist only for
$0<r<\half$, we expect that $y$ is always smaller than $1\pm 2r$, namely
that pairing fluctuations are always strongest in the quantum critical regime,
and that as
$r\rightarrow\half$,
$y$ and $1-2r$ both approach 0 before CR1 and CR2 merge with the Kondo-singlet
fixed point and disappear. We therefore conclude that the underlying
Kondo-destruction QCP promotes singlet superconducting pairing.

To summarize, we have found a quantum critical point in the two-impurity Anderson model 
with a pseudogap density of states. It
exhibits
critical Kondo destruction and shows all the hallmarks of an interacting fixed point,
such as hyperscaling relations among critical exponents and $\omega/T$ scaling in
the dynamical properties. The singlet pairing susceptibility
is found to be sharply peaked at the quantum critical point.
Our results suggest that Kondo-destruction quantum criticality promotes spin-singlet
unconventional superconductivity in a robust way and, as such, is a viable mechanism
for understanding superconductivity 
in CeRhIn$_5$ and related 
quantum critical heavy-fermion systems.

\emph{Acknowledgements.} 
We acknowledge useful discussions with 
Y.\ Chou, S.\ Kirchner, E.\  M.\ Nica,  Z.\ Wang, and H.\ Xie, as well as
technical support from the Center for Research Computing at Rice University.
This work was supported in part by NSF Grant No.\ 
DMR-1920740 and
the Robert A.\ Welch Foundation Grant No.\ C-1411 (A.C. and Q.S.), 
by JQI-NSF-PFC, LPS-MPO-CMTC and Microsoft Q (J.H.P.),
and by NSF Grant
No.\ DMR-1508122 (K.I.). 
The computation
was 
performed 
through
the Extreme Science and Engineering Discovery Environment (XSEDE) under 
the support of the NSF Grant No. DMR170109,
and 
on the Shared Computing Infrastructure
funded by NSF under grant OCI-0959097, NIH award NCRR S10RR02950, and an
IBM Shared University Research (SUR) Award
in partnership with CISCO, Qlogic and Adaptive Computing, and Rice University.

\appendix
\section{Methods}
\label{sec:appendix1}

The CT-QMC hybridization expansion algorithm allows us to stochastically sample the perturbation series in the hybridization term free of any sign problem in the infinite separation limit.
The average perturbation order exceeds $10^{3}$ per orbital for the largest inverse temperature, $\beta=3200$ at $\Gamma_{0}=1.5$.
Within our specific case, we find the auto-correlation time measured in terms of successful updates will grow not only as temperature is lowered and perturbation order increases, but also as one increases $I_{z}$ deep into the magnetic ordered phase, where a domain wall like structure can form in the imaginary time direction. Therefore, we have introduced an additional global update in addition to the standard local one kink (a kink refers to a creation and annihilation operator pair) and two kinks update, by exchanging all the kinks between different orbitals within a imaginary time interval of length around $\beta /2$ (with a probability that satisfies detailed balance), to prevent the sampling process from getting trapped in some meta-stable state.

The NRG runs were performed for Wilson discretization parameter $\Lambda=9$, retaining
between 1000 and 4000 many-body eigenstates after each iteration. The Wilsonian
discretization of the conduction band reduces the effective density of states so it
is appropriate to compare NRG calculations for hybridization width $\Gamma_0$ with
continuum-limt ($\Lambda\to 1$) results for hybridization width
$\Gamma_0 / A(\Lambda,r)$, where $A(\Lambda,r)$ is defined in Ref.\ 21 
of the main text.
Hybridization widths reported in the text are the values entered into the NRG calculations
and do not include the discretization correction factor.

\section{Finite size scaling of Binder cumulant}
\label{sec:appendix2}
\FloatBarrier

The value of $\nu^{-1}$ and $I_{c}$ is determined through minimization of the quality function $S(I_{c}, \nu^{-1})$, which is essentially the mean square deviation of the scaled data points with respect to the unknown universal function. For $k$ sets of data points represented by $\{x_{ij},y_{ij} \}$, where $i=1,\cdots,k$ labels different $\beta$ and $j$ labels different $I_{z}$, we define $S(I_{c}, \nu^{-1})=1/N \sum_{i,j} (y_{ij}-Y_{ij})^{2}$. Here $Y_{ij}$ is the estimated value of the universal function at $x_{ij}$ by linear interpolation from the rest of sets $\{x_{i^{\prime}j},y_{i^{\prime}j} \}$, $i^{\prime}\neq i$. During the scaling collapse, we start by including all the sets, and then gradually excluding the highest temperature data until the result reaches convergence. Only data points satisfying
 $ \beta^{1/\nu} (I_{z}-I_{c})/I_{c} \lesssim 1 $ are included. The best estimate of $I_{c}$ and $\nu^{-1}$ 
 is where $S(I_{c}, \nu^{-1})$ reaches its minimum $S_{min}$. 
 We estimate the error by requiring $S(I_{c}+\delta I_{c}, \nu^{-1}+ \delta \nu^{-1}) - S_{min} 
 \simeq S_{min}/2 $.

\begin{figure}[h!]
\captionsetup[subfigure]{labelformat=empty}
  \centering
  \mbox{\subfloat[]{\label{}\includegraphics[width=0.75\columnwidth]{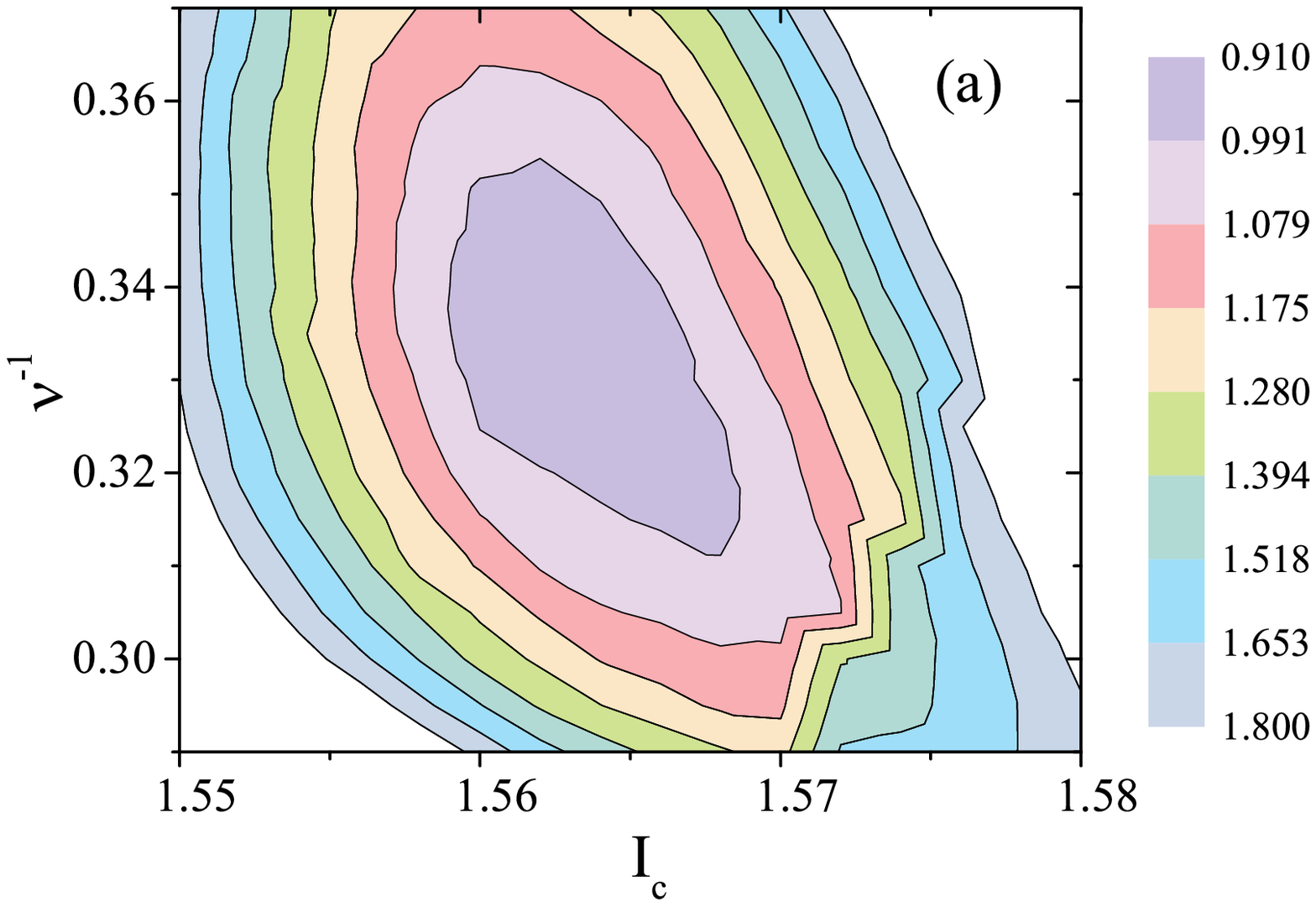}}} \\[-4ex]
   \centering
  \mbox{\subfloat[]{\label{}\includegraphics[width=0.75\columnwidth]{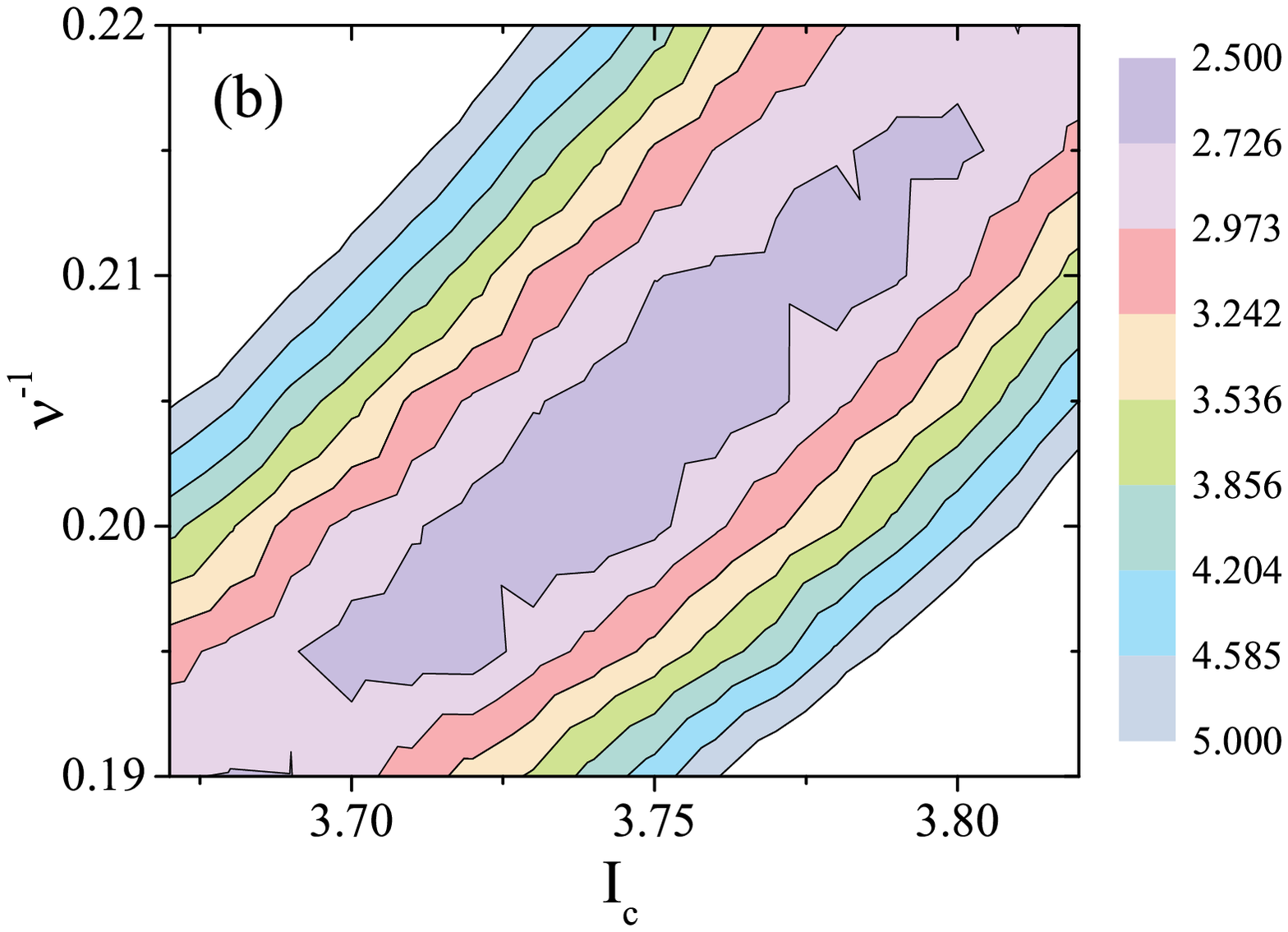}}}
\\[-2ex]
\caption{\label{fig:quality}
Contour plot of quality function $S(I_{c}, \nu^{-1}) \times 10^{4}$ for the scaling
collapse shown in Fig.\ \ref{fig:2}(c) for (a) $r=0.2$ and (b) $r=0.4$.}
\end{figure}
\FloatBarrier

\section{Additional data from CT-QMC}
\label{sec:appendix3}

\begin{figure}[h!]
\captionsetup[subfigure]{labelformat=empty}
  \centering
  \mbox{\subfloat[]{\label{}\includegraphics[width=0.65\columnwidth]{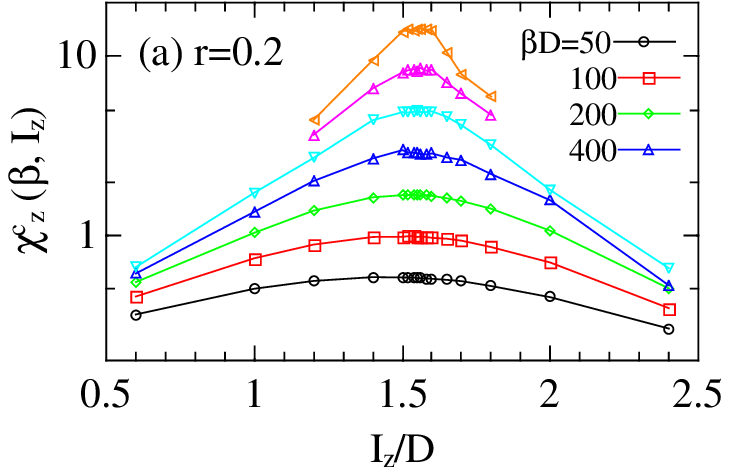}}} \\[-4ex]
    \centering                
  \mbox{\subfloat[]{\label{}\includegraphics[width=0.65\columnwidth]{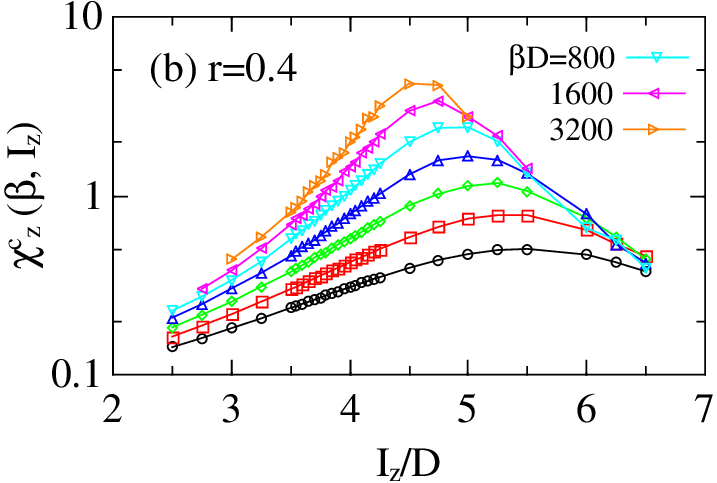}}}\\[-4ex]
   \centering
  \mbox{\subfloat[]{\label{}\includegraphics[width=0.65\columnwidth]{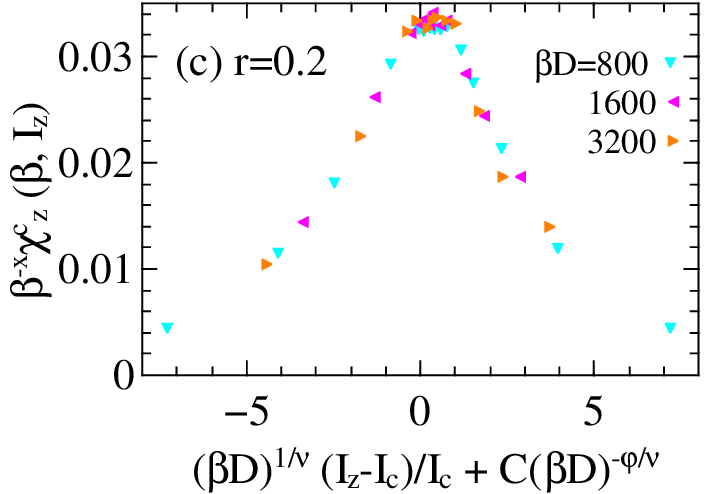}}} \\[-4ex]
    \centering                
  \mbox{\subfloat[]{\label{}\includegraphics[width=0.65\columnwidth]{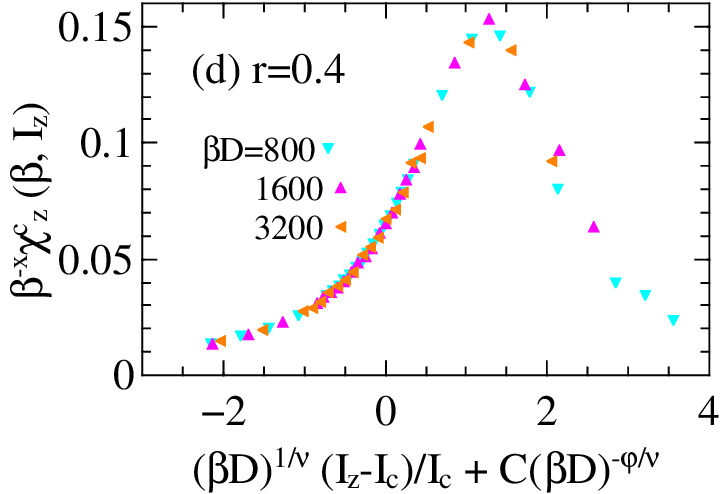}}}\\[-4ex]
\caption{\label{fig:connected}
Connected static staggered spin susceptibility $\chi_{z}^{c}(\beta,I_{z})$ vs $I_{z}$
at various inverse temperatures $\beta$ for (a) $r=0.2$, $\Gamma_{0}=0.5$ and (b) $r=0.4$,
$\Gamma_{0}=1.5$. (c)(d) Scaling collapse of $\chi_{z}^{c}(\beta,I_{z})$ in (a) and (b) respectively, with
$I_{c}=1.53(6)$, $\nu^{-1}=0.37(8)$, $x=0.75(4)$ at $r=0.2$ and $I_{c}=4.0(2)$,
$\nu^{-1}=0.26(4)$, $x=0.42(7)$ at $r=0.4$. 
The deviation from exponents in Table \ref{table:1} can be attributed to stronger finite-size
corrections to $\chi_{z}^{c}$.}
\end{figure}

\begin{figure}[h!]
\captionsetup[subfigure]{labelformat=empty}
  \centering
  \mbox{\subfloat[]{\label{}\includegraphics[width=0.65\columnwidth]{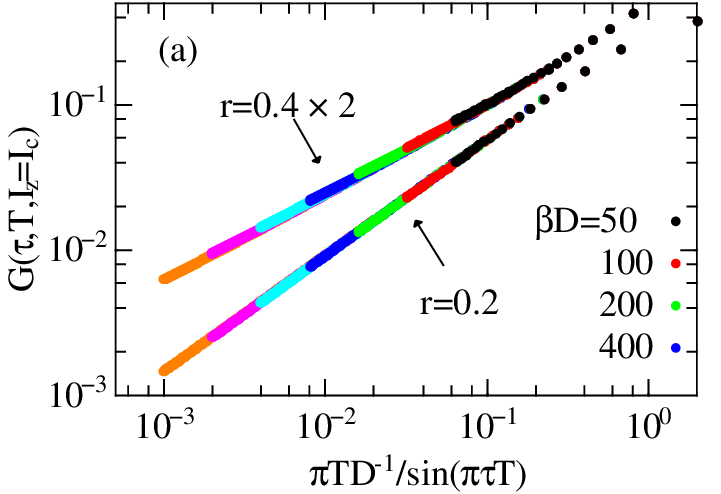}}}
  \\[-4ex]
   \centering
  \mbox{\subfloat[]{\label{}\includegraphics[width=0.65\columnwidth]{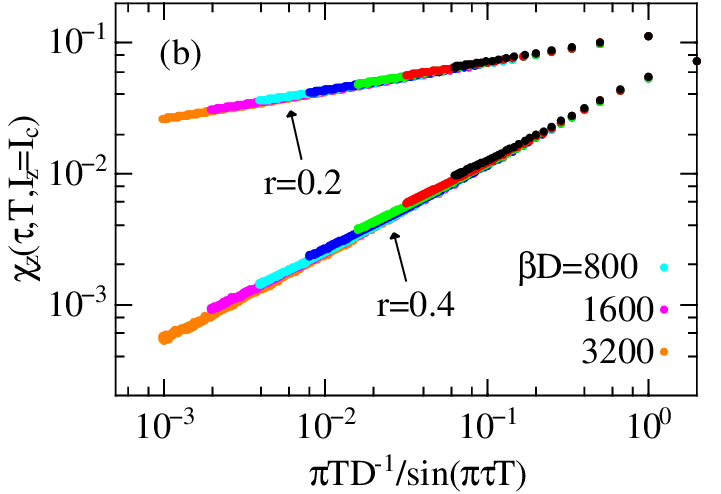}}}
\\[-4ex]
\caption{\label{fig:G_chi_critical}
(a) Scaling of the single-particle Green's function $G(\tau, T,I_{z}=I_{c})$
(averaged over impurity site and spin) with $\pi T/ \sin (\pi \tau T)$.
We find $G(\tau\rightarrow \infty, T\rightarrow 0,I_{z}=I_{c})\sim[\pi T/ \sin (\pi \tau T)]^{\eta_{G}(r)}$
with $\eta_{G}(0.2)=0.795$ and $\eta_{G}(0.4)=0.600$, consistent with $\eta_{G}=1-r$
(b) Scaling of the staggered spin correlation function $\chi_{z}(\tau, T,I_{z}=I_{c})$
with $\pi T/ \sin (\pi \tau T)$. We find
$\chi_{z}(\tau\rightarrow \infty, T\rightarrow 0,I_{z}=I_{c})\sim[\pi T/ \sin (\pi \tau T)]^{\eta_{\chi}(r)}$
with $\eta_{\chi}(0.2)=0.213$ and $\eta_{\chi}(0.4)=0.657$, consistent with $\eta_{\chi}=1-x$.
Calculations are performed at $\Gamma_{0}=0.5$ at $r=0.2$ and $\Gamma_{0}=1.5$ at $r=0.4$. }
\end{figure}

\begin{figure}[h!]
\captionsetup[subfigure]{labelformat=empty}
  \centering
  \mbox{\subfloat[]{\label{}\includegraphics[width=0.65\columnwidth]{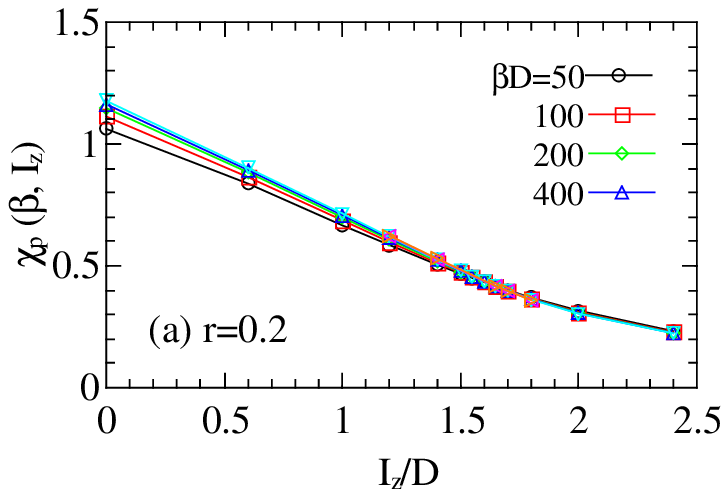}}}
\\[-4ex]
    \centering                
  \mbox{\subfloat[]{\label{}\includegraphics[width=0.65\columnwidth]{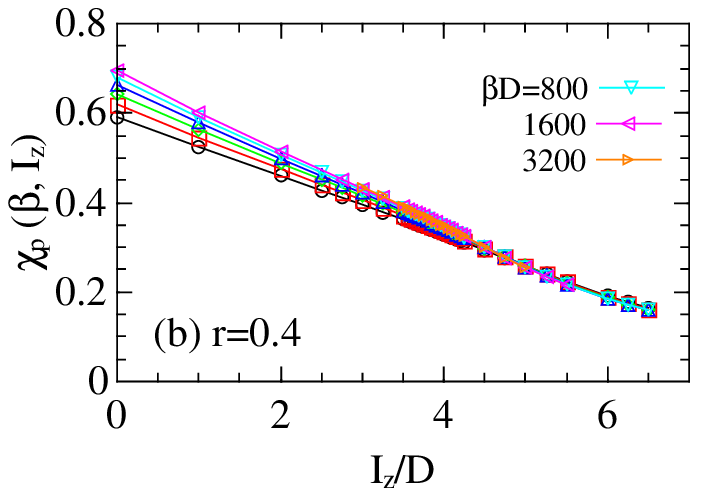}}}
\\[-4ex]
\caption{\label{fig:triplet}
Static triplet pairing susceptibility $\chi_{p}(\beta,I_{z})$ vs $I_{z}$ at various inverse
temperatures $\beta$ for (a) $r=0.2$, $\Gamma_{0}=0.5$ and (b) $r=0.4$, $\Gamma_{0}=1.5$.}
\end{figure}

\FloatBarrier

\begin{figure}[h!]
\captionsetup[subfigure]{labelformat=empty}
  \centering
  \mbox{\subfloat[]{\label{}\includegraphics[width=0.65\columnwidth]{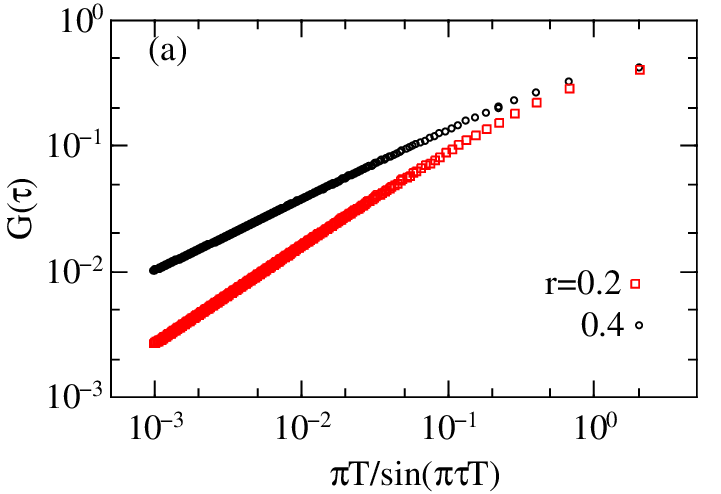}}}
  \\[-4ex]
   \centering
  \mbox{\subfloat[]{\label{}\includegraphics[width=0.65\columnwidth]{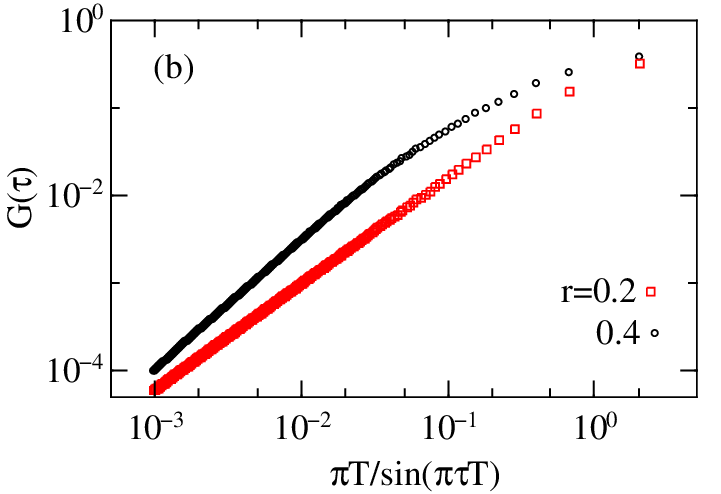}}}
\\[-4ex]
\caption{\label{fig:G_tau_Kondo_LM}
Single-impurity Green's function $G(\tau)$ at $\beta=3200$ in
(a) the Kondo-screened phase ($I_{z}=0$, $\Gamma_{0}=0.5$ for both $r=0.2$ and $r=0.4$)
and (b) the local-moment phase ($I_{z}=3$, $\Gamma_{0}=0.5$ for $r=0.2$ and $I_{z}=2$,
$\Gamma_{0}=0.5$ for $r=0.4$). Fitting to
$G_{i,\sigma}(\tau) \sim[\pi T/ \sin (\pi \tau T)]^{\eta_{G}(r)}$ in (a) gives
$\eta_{G}(0.2)=0.77$ and $\eta_{G}(0.4)=0.57$, while in (b) gives
$\eta_{G}(0.2)=1.21$ and $\eta_{G}(0.4)=1.45$. Results are consistent with
$G(\tau) \sim 1/\tau^{1\pm r}$ for $I_{z}>I_{c}$ or $I_{z}<I_{c}$.}
\end{figure}

\begin{figure}[h!]
\captionsetup[subfigure]{labelformat=empty}
  \centering
  \mbox{\subfloat[]{\label{}\includegraphics[width=0.65\columnwidth]{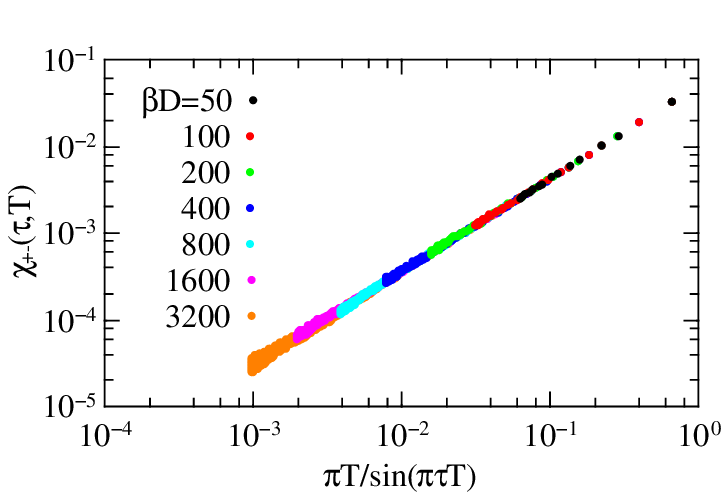}}}
\\[-2ex]
\caption{\label{fig:chi_transverse}
Transverse component of the staggered spin susceptibility $\chi_{+-}(\tau)$
at $I_{z}=I_{c}$, $\Gamma_{0}=0.5$ for $r=0.2$.
$\chi_{+-}(\tau) =  \int_{0}^{\beta} d\tau  \langle T_{\tau} \frac{1}{2} (S_{1}^{+}(\tau)-S_{2}^{+}(\tau))
\frac{1}{2} (S_{1}^{-}-S_{2}^{-}) \rangle /2 $, with $S_{i}^{-}= d_{i \dn}^{\dag}d_{i \up}$ and
$S_{i}^{+}= {S_{i}^{-}}^{\dag}$, such that $\chi_{z}=\chi_{+-}$ at the $I_{z}=0$ SU(2)-symmetric
point. The result is consistent with $\chi_{+-}(\tau)\sim 1/ \tau^{1+y}$, with $y$ taking the same
value (within numerical uncertainty) as found in the singlet pairing susceptibility $\chi_{d}(\tau)$.}
\end{figure}

\section{Additional data from NRG}
\label{sec:appendix4}
\FloatBarrier

\begin{figure}[h!]
\captionsetup[subfigure]{labelformat=empty}
  \centering
  \mbox{\subfloat[]{\label{}\includegraphics[width=0.65\columnwidth]{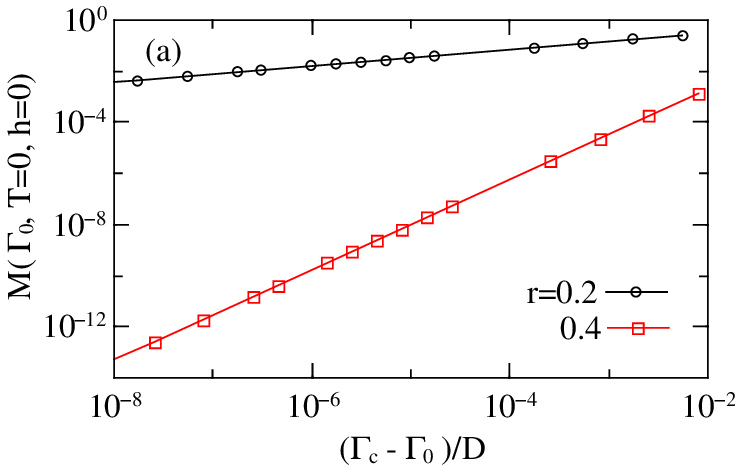}}}
  \\[-4ex]
   \centering
  \mbox{\subfloat[]{\label{}\includegraphics[width=0.65\columnwidth]{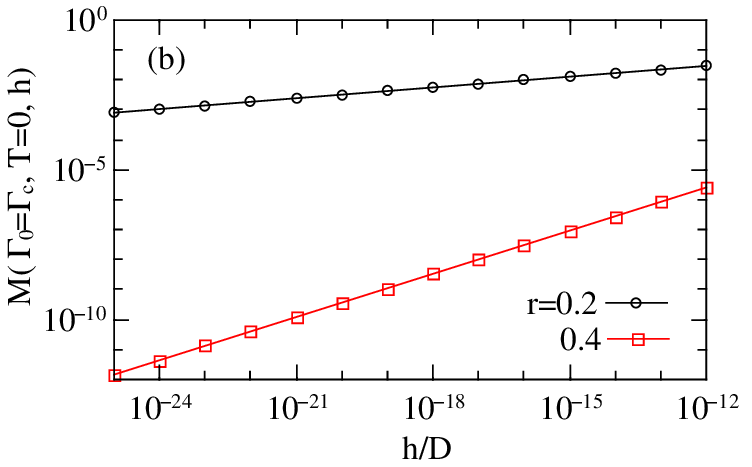}}}
\\[-4ex]
\caption{\label{fig:beta_delta}
(a) Staggered local moment $M(\Gamma_{0},T=0,h=0)$ vs $\Gamma_{c}-\Gamma_{0}$,
fitted to $M \propto (\Gamma_{c}-\Gamma_{0})^{\beta'} $ with $\beta'=0.31991(2)$ for $r=0.2$
and $\beta'=1.7701(2)$ for $r=0.4$. (b) Staggered local moment $M(\Gamma_{0}=\Gamma_{c},T=0,h)$
vs staggered external magnetic field $h$, fitted to $M \propto h ^{1/\delta} $ with
$1/\delta=0.11990(4)$ for $r=0.2$ and $1/\delta=0.48066(4)$ for $r=0.4$. Calculations are
performed at $I_{z}=1.54\:(0.73)$ for $r=0.2\:(0.4)$.}
\end{figure}

\begin{figure}[h!]
\captionsetup[subfigure]{labelformat=empty}
  \centering
  \mbox{\subfloat[]{\label{}\includegraphics[width=0.65\columnwidth]{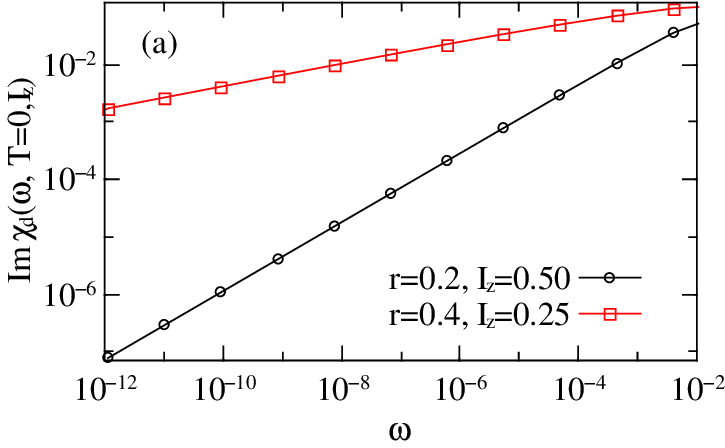}}}
  \\[-4ex]
   \centering
  \mbox{\subfloat[]{\label{}\includegraphics[width=0.65\columnwidth]{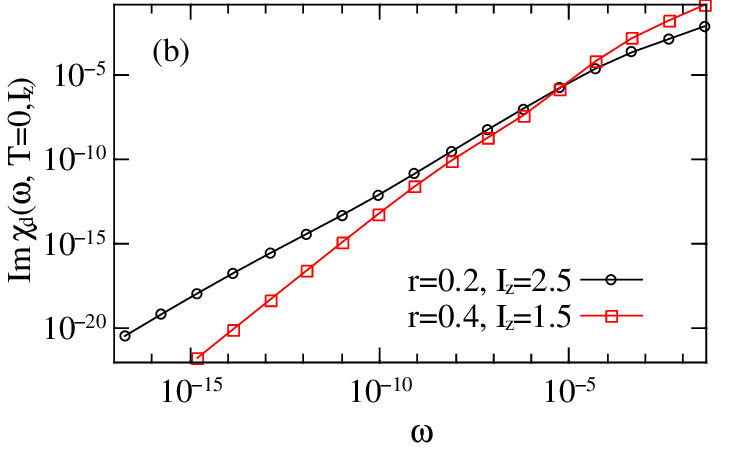}}}
\\[-4ex]
\caption{\label{fig:Im_chi_Kondo_LM}
$\operatorname{Im} \chi_d(\omega)$ at $T=0$ in (a) Kondo-screened and (b) local-moment
phases. The low-frequency asymptotics give
(a) $\operatorname{Im} \chi_d(\omega) \sgn(\omega) \propto|{\omega}|^{1-2r}$ and
(b) $\operatorname{Im} \chi_d(\omega) \sgn(\omega) \propto|{\omega}|^{1+2r}$.
Calculations are performed at $\Gamma_{0}=0.5503$ and $\Gamma_{0}=0.8032$, respectively.}
\end{figure}

\FloatBarrier


\section{Derivation of Eq. (8)}
\label{sec:appendix5}
We make use of the Kramers-Kronig relation
\begin{equation}
 \operatorname{Re} \chi_{d} (\omega=0) = \frac{1}{\pi} \int_{-\infty}^{\infty} d \omega^{\prime}  \frac{\operatorname{Im} \chi_{d} (\omega^{\prime})}{\omega^{\prime}}.
\end{equation}
Because $\operatorname{Im} \chi_{d}(\omega^{\prime})$ is odd,
\begin{eqnarray}
&& \operatorname{Re} \chi_{d} (\omega=0) \nonumber  \\
 &=& \frac{2}{\pi} \int_{0}^{\infty} d \omega  \frac{\operatorname{Im} \chi_{d} (\omega)}{\omega} \nonumber \\
 &=& \frac{2}{\pi} (\int_{0}^{\omega^{*}} d \omega+\int_{\omega^{*} }^{ \omega_{1}} d \omega +\int_{\omega_{1} }^{ \infty } d \omega )  \frac{\operatorname{Im} \chi_{d} (\omega)}{\omega},
\end{eqnarray}
where $\omega^{*} \sim |(I_{z}-I_{c})/I_{c})|^{\nu}$ is the crossover scale into the quantum critical regime, and $\omega_{1}$ is some upper cutoff, which we have assumed to be independent of $I_{z}$. The high-frequency non-universal part should only have a weak dependence on $I_{z}$, so we put $\int_{\omega_{1} }^{ \infty } d \omega \, [\operatorname{Im} \chi_{d} (\omega)] / \omega \simeq D_{1}(r)$.

Now we substitute Eq.\ \eqref{pair_freq} in the main text. 
\begin{eqnarray}
 &&\operatorname{Re} \chi_{d} (\omega=0) \nonumber \\
 &=&  \frac{2}{\pi}  \int_{0}^{\omega^{*}}   C \left(\frac{\omega^{*}}{D}\right)^{y} \left(\frac{\omega}{\omega^{*}}\right)^{1\pm 2r} \frac{1}{\omega} d \omega \nonumber \\
& +&  \frac{2}{\pi}   \int_{\omega }^{ \omega_{1}}   C \left(\frac{\omega^{*}}{D}\right)^{y}  \frac{1}{\omega} d \omega
 +D_{1}(r) \nonumber \\
 &=&  \frac{2}{\pi} C \left(\frac{\omega^{*}}{D}\right)^{y} \frac{1}{1\pm 2r}
 + \frac{2}{\pi} C \frac{1}{y} \frac{( \omega_{1}^{y} - \omega^{*y})}{ D^{y} }
  +D_{1}(r)\nonumber \\
  &=& \frac{2C}{\pi} \left[\frac{1}{y} \frac{\omega_{1}^{y}}{D^{y}}- \left(\frac{\omega^{*}}{D}\right)^{y} \left(\frac{1}{y}-\frac{1}{1\pm 2r}\right) \right]
  +D_{1}(r)
\end{eqnarray}
From NRG data, the proportionality constant $C$ has negligible dependence on $I_{z}$. Finally we replace $\omega^{*}$ by $|(I_{z}-I_{c})/I_{c})|^{\nu}$ to obtain
\begin{eqnarray}
&&\operatorname{Re} \chi_{d}(\omega=0) \nonumber \\
 &=& C_{1}(r) \nonumber \\
 &-& C_{2}(r) \left( \frac{1}{y}- \frac{1}{ 1\pm 2r } \right) \left( \frac{|(I_{z}-I_{c})/I_{z}|^{\nu}}{D} \right) ^{y},
\end{eqnarray}
where
\begin{eqnarray}
C_{1}(r)= \frac{2C(r)}{\pi} \frac{1}{y} \frac{\omega_{1}^{y}}{D^{y}} + D_{1}(r), \quad C_{2}(r)=\frac{2C(r)}{\pi} .
\end{eqnarray}


\begin{thebibliography}{31}
\expandafter\ifx\csname natexlab\endcsname\relax\def\natexlab#1{#1}\fi
\expandafter\ifx\csname bibnamefont\endcsname\relax
  \def\bibnamefont#1{#1}\fi
\expandafter\ifx\csname bibfnamefont\endcsname\relax
  \def\bibfnamefont#1{#1}\fi
\expandafter\ifx\csname citenamefont\endcsname\relax
  \def\citenamefont#1{#1}\fi
\expandafter\ifx\csname url\endcsname\relax
  \def\url#1{\texttt{#1}}\fi
\expandafter\ifx\csname urlprefix\endcsname\relax\def\urlprefix{URL }\fi
\providecommand{\bibinfo}[2]{#2}
\providecommand{\eprint}[2][]{\url{#2}}

\bibitem{LeeNagaosaWen}
P. A. Lee, N. Nagaosa, and X.-G. Wen, 
Rev. Mod. Phys. \textbf{78}, 17 (2006).

\bibitem{Si2016}
Q. Si, R. Yu, and E. Abrahams, 
Nat. Rev. Mater. \textbf{1}, 16017 (2016).

\bibitem{Takabayashi2009}
Y. Takabayashi et al.,
Science \textbf{323}, 1585 (2009).

\bibitem{SiSteglich_Science2010}
Q.\ Si and F.\ Steglich, Science \textbf{329}, 1161 (2010).

\bibitem{Lohneysen_RMP2007}
H.\ L{\"o}hneysen, A.\ Rosch, M.\ Vojta, and P.\ W{\"o}lfle 
Rev.\ Mod.\ Phys.\ \textbf{79}, 1015, (2007).

\bibitem{Hertz}
J.\ A.\ Hertz, Phys.\ Rev.\ B \textbf{14}, 1165 (1976).

\bibitem{Millis}
A.\ J.\ Millis, Phys.\ Rev.\ B \textbf{48}, 7183  (1993).

\bibitem{Moriya}  T.\ Moriya
\emph{Spin Fluctuations in Itinerant Electron Magnetism},
\textbf{56}, 44--81 (Springer, 1985).

\bibitem{Si-2001}
Q.\ Si, S.\ Rabello, K.\ Ingersent, and J.\ L.\ Smith,
Nature \textbf{413}, 804 (2001).
  
\bibitem{Coleman-2001} 
P.\ Coleman, C.\ P\'epin, Q.\ Si, and R.\ Ramazashvili,
J.\ Phys.: Condens.\ Matter \textbf{13}, R723 (2001). 

\bibitem{Park_Nature2006} 
T.\ Park, F.\ Ronning, H.\ Q.\ Yuan, M.\ B.\ Salamon,
R.\ Movshovich, J.\ L.\ Sarrao, and J.\ D.\ Thompson,
Nature \textbf{440}, 65 (2006).

\bibitem{Park_2008}
T.\ Park, E.\ D.\ Bauer, and J.\ D.\ Thompson 
Phys.\ Rev.\ Lett.\ \textbf{101}, 177002 (2008).

\bibitem{Qimiao_JPSJ_1} O.\ Stockert, S.\ Kirchner, F .\ Steglich, Q.\ Si, 
J.\ Phys.\ Soc.\ Jpn.\ \textbf{81}, 011001 (2012).

\bibitem{Qimiao_JPSJ_2} Q.\ Si, J.\ H.\ Pixley, E.\ Nica,
S.\ J.\ Yamamoto, P.\ Goswami,  R.\ Yu, and S.\ Kirchner, 
J.\ Phys.\ Soc.\ Jpn.\ \textbf{83}, 061005 (2014).

\bibitem{Shishido.2005}
H.\ Shishido, R.\ Settai, H.\ Harima, and Y.\ {\=O}nuki,
J.\ Phys.\ Soc.\ Jpn.\ \textbf{74}, 1103 (2005).

\bibitem[{\citenamefont{Pixley et~al.}(2013{\natexlab{a}})\citenamefont{Pixley,
  Kirchner, Ingersent, and Si}}]{pixley2013quantum}
\bibinfo{author}{\bibfnamefont{J.~H.}~\bibnamefont{Pixley}},
  \bibinfo{author}{\bibfnamefont{S.}~\bibnamefont{Kirchner}},
  \bibinfo{author}{\bibfnamefont{K.}~\bibnamefont{Ingersent}},
  \bibnamefont{and} \bibinfo{author}{\bibfnamefont{Q.}~\bibnamefont{Si}},
  \bibinfo{journal}{Phys.\ Rev.\ B} \textbf{\bibinfo{volume}{88}},
  \bibinfo{pages}{245111} (\bibinfo{year}{2013}{\natexlab{a}}).

\bibitem[{\citenamefont{Kirchner and Si}(2008)}]{kirchner2008scaling}
\bibinfo{author}{\bibfnamefont{S.}~\bibnamefont{Kirchner}} \bibnamefont{and}
  \bibinfo{author}{\bibfnamefont{Q.}~\bibnamefont{Si}},
  \bibinfo{journal}{Phys.\ Rev.\ Lett.\ } \textbf{\bibinfo{volume}{100}},
  \bibinfo{pages}{026403} (\bibinfo{year}{2008}).

\bibitem[{\citenamefont{Pixley et~al.}(2012)\citenamefont{Pixley, Kirchner,
  Ingersent, and Si}}]{pixley2012kondo}
\bibinfo{author}{\bibfnamefont{J.~H.}~\bibnamefont{Pixley}},
  \bibinfo{author}{\bibfnamefont{S.}~\bibnamefont{Kirchner}},
  \bibinfo{author}{\bibfnamefont{K.}~\bibnamefont{Ingersent}},
  \bibnamefont{and} \bibinfo{author}{\bibfnamefont{Q.}~\bibnamefont{Si}},
  \bibinfo{journal}{Phys.\ Rev.\ Lett.\ } \textbf{\bibinfo{volume}{109}},
  \bibinfo{pages}{086403} (\bibinfo{year}{2012}).

\bibitem{Ingersent.Si.2002}
K.\ Ingersent and Q.\ Si,
Phys.\ Rev.\ Lett.\ \textbf{89}, 076403 (2002).

\bibitem{Glossop.Kirchner.Pixley.Si.2011}
M.\ T.\ Glossop, S.\ Kirchner, J.\ H.\ Pixley, and Q.\ Si,
Phys.\ Rev.\ Lett.\ \textbf{107}, 076404 (2011).

\bibitem{Formalism}
J.\ H.\ Pixley, A.\ Cai and Q.\ Si,
Phys.\ Rev.\ B \textbf{91}, 125127 (2015).

\bibitem{entanglement}
J.\ H.\ Pixley, T.\ Chowdhury, M.\ T.\ Miecnikowski, J.\ Stephens, C.\ Wagner, and K.\ Ingersent,
Phys.\ Rev.\ B \textbf{91}, 245122 (2015).

\bibitem{Pixley.2015}
J.\ H.\ Pixley, L.\ Deng, K.\ Ingersent, and Q.\ Si,
Phys.\ Rev.\ B \textbf{91}, 201109(R) (2015).

\bibitem{fradkin}
D.\ Withoff and E.\ Fradkin
Phys.\ Rev.\ Lett.\ \textbf{64}, 1835 (1990).

\bibitem{Bulla.1997}
R.\ Bulla, T.\ Pruschke, and A.\ C.\ Hewson,
J.\ Phys.: Condens.\ Matter \textbf{9}, 10463 (1997).

\bibitem{Buxton.1998}
C.\ Gonzalez-Buxton, and K.\ Ingersent,
Phys.\ Rev.\ B \textbf{57}, 14254 (1998).

\bibitem{Garst.etal}
In the absence of the pseudogap, our model exhibits a Kosterlitz-Thouless
quantum phase transition between Kondo and Kondo-destroyed phases; see
M.\ Garst, S.\ Kehrein, T.\ Pruschke, A.\ Rosch, and M.\ Vojta,
Phys.\ Rev.\ B \textbf{69}, 214413 (2004), also
N.\ Andrei, G.\ T.\ Zim{\'a}nyi, G.\ Sch{\"o}n, Phys. Rev. B {\bf 60}, R5125 (1999).

\bibitem{Affleck.Ludwig}
I.\ Affleck and A.\ W.\ W.\ Ludwig, Phys. Rev. Lett. {\bf 68}, 1046 (1992); I.\ Affleck, A.\ W.\ W.\ Ludwig, and B.\ A.\ Jones, Phys. Rev. B {\bf 52}, 9528 (1995).

\bibitem{Gan}
J.\ Gan, Phys. Rev. Lett. {\bf 74}, 2583 (1995); Phys. Rev. B {\bf 51}, 8287 (1995).

\bibitem{Zarand}
G.\ Zar{\'a}nd, C.\ Chung,\ P.\ Simon, and M.\ Vojta, Phys. Rev. Lett. {\bf 97}, 166802 (2006).

\bibitem[{\citenamefont{Werner and Millis}(2007)}]{Werner-2007}
\bibinfo{author}{\bibfnamefont{P.}~\bibnamefont{Werner}} \bibnamefont{and}
  \bibinfo{author}{\bibfnamefont{A.~J.} \bibnamefont{Millis}},
  \bibinfo{journal}{Phys. Rev. Lett.} \textbf{\bibinfo{volume}{99}},
  \bibinfo{pages}{146404} (\bibinfo{year}{2007}).

\bibitem[{\citenamefont{Werner and Millis}(2010)}]{Werner-2010}
\bibinfo{author}{\bibfnamefont{P.}~\bibnamefont{Werner}} \bibnamefont{and}
  \bibinfo{author}{\bibfnamefont{A.~J.} \bibnamefont{Millis}},
  \bibinfo{journal}{Phys. Rev. Lett.} \textbf{\bibinfo{volume}{104}},
  \bibinfo{pages}{146401} (\bibinfo{year}{2010}).

\bibitem{Wilson.1975}
K.\ G.\ Wilson, Rev.\ Mod.\ Phys.\ \textbf{47}, 773 (1975).

\bibitem{Bulla.2008}
R.\ Bulla, T.\ A.\ Costi, and T.\ Pruschke,
Rev.\ Mod.\ Phys.\ \textbf{80}, 395 (2008).

\bibitem{Binder}
K.\ Binder, Z.\ Phys.\ B \textbf{43}, 119 (1981).

\bibitem{Houdayer.Hartmann.2004}
J.\ Houdayer and A.\ K.\ Hartmann,
Phys.\ Rev.\ B \textbf{70}, 014418 (2004).

\bibitem{correction_to_scaling}
For $r=0.4$, sub-leading corrections to scaling are sufficiently strong
that it proved challenging for CT-QMC to reproduce the NRG value of $\nu^{-1}$
within estimated error. We attained this goal by using NRG calculations to
select the value $\Gamma_0=1.5$ that produces the highest temperature of entry
into the quantum critical regime.

\bibitem{Kircan}
M.\ Kir{\'c}an, and M.\ Vojta,
Phys.\ Rev.\ B \textbf{69}, 174421 (2004).

\bibitem{footnote}
We note that in the infinite-separation limit, the pairing correlation
of $d_{2\dn}d_{1\up} + d_{2\up}d_{1\dn}$ is degenerate with that of
$d_{2\dn}d_{1\up} - d_{2\up}d_{1\dn}$.

\bibitem{performance_analysis}
E.\ Gull, P.\ Werner, A.\ Millis, and M.\ Troyer, Phys. Rev. B \textbf{76}, 235123 (2007).
\end{thebibliography}
\end{document}